# A Hebbian/Anti-Hebbian Network Derived from Online Non-Negative Matrix Factorization Can Cluster and Discover Sparse Features

Cengiz Pehlevan[1,2] and Dmitri B. Chklovskii[2]

[1] Janelia Farm Research Campus
Howard Hughes Medical Institute
Ashburn, VA 20147
pehlevanc@janelia.hhmi.org

[2] Simons Center for Data Analysis
Simons Foundation
New York, NY 10010
mitya@simonsfoundation.org

*Abstract:* Despite our extensive knowledge of biophysical properties of neurons, there is no commonly accepted algorithmic theory of neuronal function. Here we explore the hypothesis that single-layer neuronal networks perform online symmetric non-negative matrix factorization (SNMF) of the similarity matrix of the streamed data. By starting with the SNMF cost function we derive an online algorithm, which can be implemented by a biologically plausible network with local learning rules. We demonstrate that such network performs soft clustering of the data as well as sparse feature discovery. The derived algorithm replicates many known aspects of sensory anatomy and biophysical properties of neurons including unipolar nature of neuronal activity and synaptic weights, local synaptic plasticity rules and the dependence of learning rate on cumulative neuronal activity. Thus, we make a step towards an algorithmic theory of neuronal function, which should facilitate large-scale neural circuit simulations and biologically inspired artificial intelligence.

## I. INTRODUCTION

Our brains are capable of performing effectively and efficiently a variety of different computations. Focusing on unsupervised tasks alone, neuronal circuits must perform, for example, clustering and feature discovery. Both of these tasks may reduce or expand the dimensionality of the input. At the same time the elementary building blocks of the brain hardware, neurons and synapses, are physiologically stereotypical. Is there a generic algorithm implementable by biologically plausible neuronal circuits, which is capable of both clustering and feature discovery?

Previously, the machine learning literature has discussed a connection between clustering and feature discovery. Specifically, *K*-means, a commonly used clustering algorithm can discover sparse features [1], [2] and perform independent component analysis (ICA) [3]. However, algorithm implementation in a biological plausible neural network has not been addressed.

Clustering by neural networks has been considered previously in a biologically relevant online setting where samples are presented sequentially one at a time and assigned to a cluster immediately upon each presentation. Examples of "Competitive Hebbian Learning" algorithms (reviewed in [4], [5]) include the incremental *K*-means algorithm [6], the self-organizing maps (SOM) [7], the neural gas [8] and the adaptive resonance theory (ART) [9], [10]. Such networks typically implement a competitive, "winner-take-all" type dynamics among neurons, where the winner neuron signals assignment of a datum to the neuron's associated cluster, followed by a Hebbian update to the synaptic weights of the winner neuron, which encode the cluster centroid. While the Hebbian update can be derived from a clustering cost function in certain cases [11], this not the case for the connectivity and dynamics implementing competition. Instead, such connectivity and dynamics is prescribed as in ART, or assumed to exist as in SOM, the neural gas, or the incremental *K*-means algorithms.

At the same time, neural networks can discover features, such as principal components [12]–[17], independent components [18]–[20], or sparse overcomplete representations [21], [22]. Single-layer implementations of such networks vary in details: some are derived from a principled cost function but rely on non-local and hence biologically implausible learning rules [16], [18], [20]–[22], others use local learning rules [13]–[15] but cannot be derived from a principled cost function. Therefore, except [17], these contributions have not been able to combine a derivation from a principled cost function with biologically plausible local learning rules.

In this paper, starting from a principled cost function, SNMF of the similarity of the input data, we derive an algorithm that can perform both online clustering and feature discovery. The algorithm maps onto a network using only biologically plausible local learning rules: Hebbian and anti-Hebbian.

To introduce our notation, the input to the algorithm is denoted by data matrix, $X$, which contains $T$ column-vectors, $x_t \in \mathbb{R}^n$ and a similarly defined nonnegative output data matrix, $Y \in \mathbb{R}_+^{m \times T}$:

$$X = (x_1 \cdots x_T) = \begin{pmatrix} x_{1,1} & \cdots & x_{T,1} \\ \vdots & & \vdots \\ x_{1,n} & \cdots & x_{T,n} \end{pmatrix}; \quad Y = (y_1 \cdots y_T) = \begin{pmatrix} y_{1,1} & \cdots & y_{T,1} \\ \vdots & & \vdots \\ y_{1,n} & \cdots & y_{T,n} \end{pmatrix}.$$

We denote the transpose of matrix $A$ as $A'$ and its Frobenius norm as $\|A\|_F$.

Our approach is based on the previous suggestion [23], [24] that clustering can be accomplished by solving the SNMF optimization problem:

$$Y^* = \arg\min_{Y \geq 0} \|X'X - Y'Y\|_F^2, \quad (1)$$

where $Y$ is the indicator matrix whose element $Y_{t,i}$ is non-zero if sample $t$ is attributed to cluster $i$. The intuition for the clustering utility of SNMF is given in Fig. 1A.

The rest of the paper is organized as follows. In the next Section we review the formal connection between $K$-means clustering and SNMF. In Section A of the Results, we derive an online algorithm to solve SNMF and hence clustering. In Section B, we illustrate our algorithm's clustering performance on a numerical example. In Section C, we demonstrate that network operation can also be viewed in the context of sparse feature discovery. In Discussion, we propose that online symmetric matrix factorization my serve as a powerful algorithmic theory of neural computation and compare our SNMF network with known biological facts.

## II. CONNECTION BETWEEN $K$-MEANS AND SNMF

Here we review the connection between the $K$-means clustering objective function and the SNMF objective function [23]. The $K$-means algorithm, Fig. 1B, clusters data by solving the following optimization problem:

$$\min_{\{k\}} \sum_{k=1}^{K} \sum_{t \in C_k} \left\| x_t - \frac{1}{n_k} \sum_{s \in C_k} x_s \right\|_2^2,$$

which can be rewritten in the following form:

$$\min_{\{k\}} \left[ -\sum_{k=1}^{m} \frac{1}{n_k} \sum_{t,s \in C_k} x_t' x_s \right]. \quad (2)$$

As mentioned in the Introduction, a clustering solution can be represented by the indicator matrix $Y$. In $K$-means clustering, $Y$ comprises $K$ $T$-dimensional scaled binary row-vectors, $y_{:,k}$:

$$Y = \begin{pmatrix} y_{:,1} \\ \vdots \\ y_{:,K} \end{pmatrix}, \quad y_{:,k} = \frac{1}{n_k^{1/2}} \left( 0, \cdots, 0, \overbrace{1, \cdots, 1}^{n_k}, 0, \cdots, 0 \right). \quad (3)$$

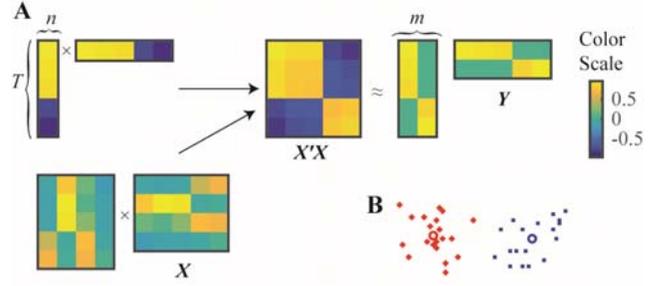

Fig. 1: Clustering and SNMF A) SNMF as a clustering tool. Matrix of inner products of the input data represents the similarity of the $T$ input samples. SNMF of the similarity matrix yields an indicator matrix, which attributes $T$ samples to $m$ clusters. Note that the dimension of the input data vectors can be less (upper example) or greater (lower example) than the number of clusters $m$. B) Schematic illustration of clustering in the space of $x$. Colors indicate cluster assignments, empty circles are cluster centroids.

Note that in $K$-means clustering, each data point belongs to one and only one cluster meaning that the rows of $Y$ are orthonormal, $YY' = I$. Let $\Pi$ be the set of all indicator matrices of type given in (3). Then, the objective function for $K$-means clustering (2) can be written as follows:

$$Y^* = \arg\min_{Y \in \Pi} \left[ -\text{Tr}(YX'XY') \right]$$

$$= \arg\min_{Y \in \Pi} \left[ -\text{Tr}(X'XY'Y) + \frac{1}{2}\text{Tr}(Y'YY'Y) \right]$$

$$= \arg\min_{Y \in \Pi} \frac{1}{2} \|X'X - Y'Y\|_F^2,$$

where the first equality follows from the orthonormality of the indicator functions and the second - by completing a square with a quartic $X$ term independent of $Y$.

Finally, by relaxing the constraint on the binary form of $Y$, but keeping it non-negative, we arrive at the SNMF cost function (1). Such relaxation permits fractional values in the indicator matrix allowing for the possibility that a given data point may belong to more than one cluster, so-called soft clustering.

## III. RESULTS

### A. Derivation of the online algorithm from the SNMF cost function

Here we minimize the SNMF cost function in the online setting where the input data are streamed sequentially one vector, $x_T$, at a time and the corresponding, $y_T$, must be computed before the arrival of $x_{T+1}$. Thus, at each time point we solve the following cost function:

$$y_T = \arg\min_{y_T \geq 0} \|X'X - Y'Y\|_F^2 + \lambda \, \text{rank}(Y). \quad (4)$$

Because in the online setting the final rank of the output may be unknown a priori, we control it by a regularizer with λ being a regularization coefficient.

To solve (4) we expand the Frobenius norm and group the terms:

$$\mathbf{y}_T = \underset{\substack{\mathbf{y}_T \geq 0 \\ \text{Card}(\mathbf{y}_T) \geq \text{Card}(\mathbf{y}_{T-1})}}{\arg\min} \left[ \sum_{t=1}^{T}\sum_{s=1}^{T}(\mathbf{x}'_t\mathbf{x}_s - \mathbf{y}'_t\mathbf{y}_s)^2 + \lambda \text{Card}(\mathbf{y}_T) \right]$$

$$= \underset{\substack{\mathbf{y}_T \geq 0 \\ \text{Card}(\mathbf{y}_T) \geq \text{Card}(\mathbf{y}_{T-1})}}{\arg\min} \left[ 2\sum_{t=1}^{T-1}(\mathbf{x}'_t\mathbf{x}_T - \mathbf{y}'_t\mathbf{y}_T)^2 + (\mathbf{x}'_T\mathbf{x}_T - \mathbf{y}'_T\mathbf{y}_T)^2 + \lambda \text{Card}(\mathbf{y}_T) \right]$$

$$= \underset{\substack{\mathbf{y}_T \geq 0 \\ \text{Card}(\mathbf{y}_T) \geq \text{Card}(\mathbf{y}_{T-1})}}{\arg\min} \sum_{i=1}^{m} \left[ \begin{array}{l} -4 y_{T,i} \sum_{j=1}^{n} x_{T,j} \sum_{t=1}^{T-1} y_{t,i} x_{t,j} \\ +2 y_{T,i} \sum_{k=1}^{m} y_{T,k} \sum_{t=1}^{T-1} y_{t,i} y_{t,k} - 2\|\mathbf{x}_T\|^2 y_{T,i}^2 \\ +y_{T,i}^2 \sum_{k=1}^{m} y_{T,k}^2 + \lambda \text{Card}(\mathbf{y}_T) \end{array} \right].$$

We solve this problem by coordinate descent on the components of $\mathbf{y}_T$:

$$y_{T,i} = \underset{\substack{y_{T,i} \geq 0 \\ \text{Card}(\mathbf{y}_T) \geq \text{Card}(\mathbf{y}_{T-1})}}{\arg\min} \left[ \begin{array}{l} -4 y_{T,i} \sum_{j=1}^{n} x_{T,j} \sum_{t=1}^{T-1} y_{t,i} x_{t,j} \\ +4 y_{T,i} \sum_{k=1, k \neq i}^{m} y_{T,k} \sum_{t=1}^{T-1} y_{t,i} y_{t,k} + 2 y_{T,i}^2 \sum_{t=1}^{T-1} y_{t,i}^2 \\ + \left( 2\sum_{k=1,k\neq i}^{m} y_{T,k}^2 - 2\|\mathbf{x}_T\|^2 \right) y_{T,i}^2 + y_{T,i}^4 \end{array} \right].$$

To enforce the cardinality constraint we consider separately the nodes that have and have not been active previously. If the i-th degree of freedom has not been utilized, i.e. $\sum_{t=1}^{T-1} y_{t,i}^2 = 0$, then

$$y_{T,i} = \underset{y_{T,i} \geq 0}{\arg\min}\left[\left(\left(\|\mathbf{x}_T\|^2 - \sum_{k=1,k\neq i}^{m} y_{T,k}^2\right) - y_{T,i}^2\right)^2 + \lambda\|y_{T,i}\|_0\right] =$$

$$= \begin{cases} 0, & \left(\|\mathbf{x}_T\|^2 - \sum_{k=1,k\neq i}^{m} y_{T,k}^2\right)^2 \leq \lambda \\ \left(\|\mathbf{x}_T\|^2 - \sum_{k=1,k\neq i}^{m} y_{T,k}^2\right)^{1/2}, & \left(\|\mathbf{x}_T\|^2 - \sum_{k=1,k\neq i}^{m} y_{T,k}^2\right)^2 > \lambda \end{cases} \quad (5)$$

Once the i-th degree of freedom has been active, i.e. $\sum_{t=1}^{T-1} y_{t,i}^2 > 0$, then

$$y_{T,i} = \underset{y_{T,i} \geq 0}{\arg\min} \sum_{t=1}^{T-1} y_{t,i}^2 \left[ \begin{array}{l} \dfrac{2\left(\sum_{k=1,k\neq i}^{m} y_{T,k}^2 - \|\mathbf{x}_T\|^2\right) y_{T,i}^2}{\sum_{t=1}^{T-1} y_{t,i}^2} \\ -4 y_{T,i} \sum_{j=1}^{n} W_{T,i,j} x_{T,j} + 2 y_{T,i}^2 + \\ 4 y_{T,i} \sum_{k=1,k\neq i}^{m} M_{T,i,k} y_{T,k} + \dfrac{y_{T,i}^4}{\sum_{t=1}^{T-1} y_{t,i}^2} \end{array} \right], \quad (6)$$

where we used the "synaptic connection" matrices defined as follows:

$$\mathbf{W}'_T = \begin{pmatrix} \mathbf{W}_{T,1} \\ \vdots \\ \mathbf{W}_{T,m} \end{pmatrix} = \begin{pmatrix} W_{T,1,1} & \cdots & W_{T,1,n} \\ \vdots & & \vdots \\ W_{T,m,1} & \cdots & W_{T,m,n} \end{pmatrix}; \quad W_{T,i,j} = \dfrac{\sum_{t=1}^{T-1} y_{t,i} x_{t,j}}{\sum_{t=1}^{T-1} y_{t,i}^2}$$

$$\mathbf{M}_T = \begin{pmatrix} \mathbf{M}_{T,1} \\ \vdots \\ \mathbf{M}_{T,m} \end{pmatrix} = \begin{pmatrix} M_{T,1,1} & \cdots & M_{T,1,m} \\ \vdots & & \vdots \\ M_{T,m,1} & \cdots & M_{T,m,m} \end{pmatrix};$$

$$M_{T,i,k\neq i} = \dfrac{\sum_{t=1}^{T-1} y_{t,i} y_{t,k}}{\sum_{t=1}^{T-1} y_{t,i}^2}; \quad M_{T,i,i} = 0$$

In the large-T limit, the first and the last terms in Eq. (6) can be ignored and the optimization reduces to:

$$y_{T,i} \approx \underset{y_{T,i} \geq 0}{\arg\min}\left(\mathbf{W}_{T,i}\mathbf{x}_T - \mathbf{M}_{T,i}\mathbf{y}_T - y_{T,i}\right)^2,$$

and, therefore,

$$y_{T,i} = \max\left(\mathbf{W}_{T,i}\mathbf{x}_T - \mathbf{M}_{T,i}\mathbf{y}_T, 0\right), \quad (7)$$

which can be naturally implemented by a neural network of summation/rectification units with feedforward and lateral synaptic connections, Fig. 2.

Synaptic weight updates for feedforward and lateral connections can be written in a recursive form, which lead to Hebbian and anti-Hebbian updates for feedforward and lateral connections respectively:

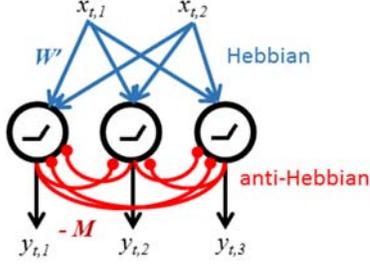

Fig. 2: A neuronal network implementing online SNMF. Each neuron rectifies the sum of inputs weighted by feedforward connections $W'$ and other neurons' activity weighted by lateral connections $-M$. Feedforward and lateral connection weights are updated according to local Hebbian and anti-Hebbian rules correspondingly.

$$\hat{Y}_{T,i} = \hat{Y}_{T-1,i} + y_{T-1,i}^2,$$
$$W_{T,i,j} = W_{T-1,i,j} + y_{T-1,i}\left(x_{T-1,j} - W_{T-1,i,j} y_{T-1,i}\right)/\hat{Y}_{T,i}, \quad (8)$$
$$M_{T,i,j\neq i} = M_{T-1,i,j} + y_{T-1,i}\left(y_{T-1,j} - M_{T-1,i,j} y_{T-1,i}\right)/\hat{Y}_{T,i},$$

Interestingly, these update rules have the functional form of Oja's rule proposed previously for single neurons [12]. Note that each weight update depends only on the activity of pre- and post-synaptic neurons, and hence is biologically plausible. Yet, unlike Oja's rule where the learning rate is arbitrary, here the learning rate is specified and activity dependent. To the best of our knowledge such single-neuron updates [25] have not been previously derived in the multi-neuron case.

Thus, starting with the SNMF cost function, we derived the following online algorithm:

**Algorithm 1: Online SNMF**

Start with the empty active set { }, $W' = 0$ and $M = 0$.

For each $T=1...T_f$

1. Initialize $y_T = 0$.
2. Receive $x_T$.
3. For each degree of freedom in the active set $\{i\}$ compute $y_{T,i} = \max\left(W_{T,i} x_T - M_{T,i} y_T, 0\right)$ iterating until convergence.
4. If required by (5) add another degree of freedom to the active set $\{i\}$.
5. Output $y_T$.
6. Update "synaptic connections" $W'$ and $M$ according to (8).

### B. Numerical test of the online clustering algorithm

Here, we demonstrate the clustering performance of our algorithm on artificial datasets. We generated a test dataset by sampling from three Gaussians (100 data points each) centered at randomly chosen locations, [(-0.0985, -0.3379), (-0.6325, 0.9322), (1.1078, 1.0856)], with identical covariance matrices $0.04 I_2$, Fig. 3A. Then, we applied Algorithm 1 with the regularization coefficient, $\lambda = 0.6$ and found that it correctly clustered the data, Fig. 3A.

We compare the performance of our online algorithm to the offline Newton-like SNMF algorithm proposed in [24]. The performance of the offline algorithm depends on initialization, and we attempted to improve it by initializing the three rows of $Y$ to i) the rectified projections of each data point onto the first principal component of the data $X$; ii) same for the second principal component; and iii) the rectified negative of the first principal component. We ran the offline algorithm on the whole dataset.

We evaluated algorithms' performance by computing the unregularized cost at time $T$ defined using the notation $X_T = (x_1,\cdots,x_T)$ for the input up to time $T$ and $Y_T = (y_1,\cdots,y_T)$ for the corresponding output:

$$C_T = \left\| X_T' X_T - Y_T' Y_T \right\|_F^2. \quad (9)$$

Here when running the online algorithm we limited the number of output degrees of freedom to three.

The ratio of the costs for the offline and the online algorithm is shown in Fig. 3B. Interestingly, we find that the performance of the online algorithm is very close to that of the offline algorithm, and, in relative terms, improves over time.

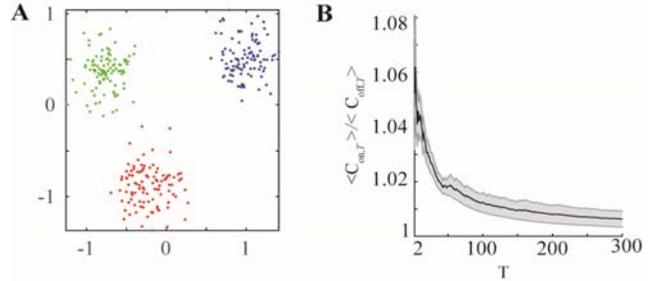

Fig. 3: Clustering of artificial datasets using online SNMF algorithm. A) An example run of our online algorithm. Dot colors indicate cluster assignments. B) Mean ratio of costs (9) of our online algorithm and of the offline algorithm [24] run on 100 datasets. $T = 1$ is not shown since the online algorithm always achieves zero cost for the initial data point. Shadows show standard deviation.

### C. Sparse feature discovery by the online SNMF algorithm

In the previous section we considered the performance of our algorithm when presented with clearly clusterable data.

However, nervous systems may encounter stimuli that do not have obvious cluster structure. In this Section, we discuss the performance of our online SNMF algorithm when presented with such natural stimuli.

We present our algorithm with a natural image ensemble as is common in computational neuroscience and computer vision. Specifically, we present sequentially $5 \times 10^4$ 16-pixel by 16-pixel natural image patches, using the data set provided in [21]. The patches are pre-processed by centering, contrast-normalization, and whitening [2], which may not be far from the computations performed in the mammalian retina and thalamus upstream of the primary visual cortex (V1) [26], [27]. Each image patch was presented to the network 40 times, in random order, resulting in a total of $2 \times 10^6$ presentations. The number of output units, $m = 256$, and the regularizer, $\lambda = 200$. To speed up convergence, we used an alternative step size schedule, where at initialization of a neuron $\hat{Y}_{T,i} = 1000$ and at subsequent steps $\hat{Y}_{T,i} = \hat{Y}_{T-1,i} + 0.01 y_{T-1,i}^2$.

The resulting feedforward weight matrix, $W'$, transformed into the neural filters acting on natural images is shown in Fig. 4A. The transformation involves right-multiplying $W'$ by the whitening matrix, $Q$, and then plotting the rows of the product. The recovered neural filters resemble Gabor-filter receptive fields of V1 neurons [28]. Previously, such Gabor-filter receptive fields were obtained as independent components of an image patch [19] as well as sparse dictionary learning model [29], where each image patch is represented by a small set of active neurons [21], [22]. Similarly, we find that the activity in our network is sparse, Fig. 4B, with a high probability of neural activity being zero. Therefore, our algorithm is capable of recovering sparse features.

Can we understand why SNMF discovers sparse features? A hint comes from the recently reported success of $K$-means clustering algorithm in discovering Gabor filters from whitened natural image patches [1], [2]. Since, as discussed in Section II, SNMF is related to $K$-means it is possible that SNMF also discovers sparse features.

Next, we review an argument showing that $K$-means discovers sparse features [3]. Let us assume that the data are generated by the Independent Component Analysis (ICA) model [19],

$$x_T = As_T, \quad (10)$$

where $A \in \mathbb{R}^{n \times n}$ is a mixing matrix, assumed to be invertible, and the random source vector, $s_T$, has statistically independent elements. The source is assumed to be sparse, e.g. Laplace distributed. Such generative model may describe natural images with rows of $A^{-1}$ corresponding to Gabor filter. When applied to data such filters would recover the original sparse sources, $A^{-1}x_T = A^{-1}As_T = s_T$ [19], [20]. Below we will assume that the number of clusters, $K = 2n$.

Central to the argument that $K$-means discovers sparse features is that it satisfies the so-called Rotation Invariance and Sparse Selectivity (RISS) property [3]:

i) Rotation Invariance: If the data are rotated by an orthonormal transformation, the cluster centroids given by the algorithm are also rotated.
ii) Sparse Selectivity: If the mixing matrix were identity, cluster centroids would be $n$ unit vectors along coordinate axis and their negatives. These directions are the sparse directions.

The RISS property allows one to prove that an algorithm discovers sparse features by the following logic. First, a RISS clustering algorithm applied to whitened input, generated from the ICA model (10), finds $2n$ cluster centroids aligned with the sparse directions. Second, a known linear transformation of such centroids can recover the rows of $A^{-1}$, or Gabor filters for natural image input. Below, we show this more formally.

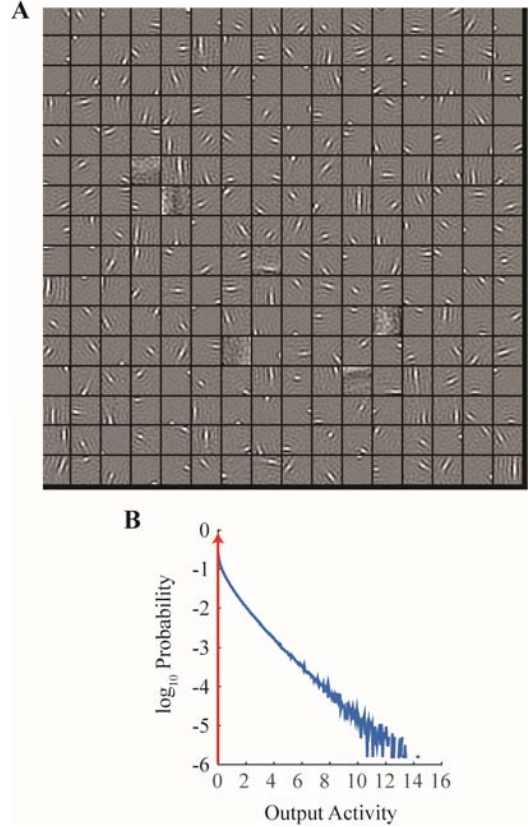

Fig. 4: SNMF network discovers sparse features in natural images. A) The neural filters trained on natural images, resemble Gabor filters. B) Probability density of output activity, pooled over all neurons and all stimuli presentations. The red arrow denotes the probability of having zero output. This plot was obtained by presenting the network with the same image patches as used for training but with frozen synaptic weights.

The data generated from the ICA model (10) and then whitened is related to the original sources by an orthogonal rotation [19]. To see this, we rewrite the whitened input, $z_T$, in

terms of the assumed to be known whitening matrix, $Q$, and by substituting (10) find:

$$z_T = Qx_T = QAs_T. \qquad (11)$$

By denoting $QA \equiv U$ we obtain from (11):

$$z_T = Us_T.$$

To show that matrix $U$ is orthonormal we start with the ortonormality of whitened data [19]:

$$I_n = \langle z_T z_T' \rangle_T = U \langle s_T s_T' \rangle_T U' = UU'.$$

Because $z_T$ is an orthogonal rotation of $s_T$ the RISS property guarantees that cluster centroids will be columns of $U$ and their negatives. Then, the ICA filters, i.e. rows of $A^{-1}$, are recovered from $U$ by reversing the whitening operation.

$$U'Q = UQAA^{-1} = U'UA^{-1} = A^{-1}.$$

Whereas the above analysis assumes that the number of clusters, $K = 2n$, this constraint is not necessary for numerically discovering Gabor-filters by the $K$-means algorithm [2], [3].

The argument summarized above can explain sparse feature discovery by the online SNMF through its relation to $K$-means algorithm, Section II. Although we were not able to prove the RISS property of our online SNMF algorithm, it satisfies the RISS property numerically, Fig. 5. If the RISS property of the online SNMF algorithm were proven, the above argument could be applied to SNMF directly.

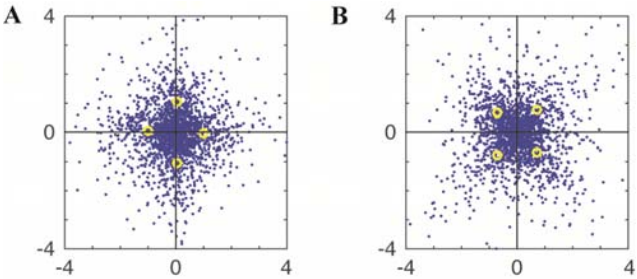

Fig. 5: Numerical verification of the RISS property for the online NMF algorithm. A) Two-dimensional input data (blue dots) is generated by sampling each coordinate from an independent Laplace distribution with unit variance. Unit vectors along positive and negative axes (yellow circles) are the rows of $W'$ generated by the online SNMF algorithm (8) in steady state. B) When input data are generated by an orthonormal mixing matrix, rows of $W'$ rotate correspondingly and point along sparse directions. For both A) and B), the maximum number of clusters was 4, and $\lambda = 4$.

## IV. DISCUSSION

### A. Computational function of online SNMF

Although we initially motivated the use of SNMF by its connection with $K$-means (Section II) we believe that its computational role maybe broader. As shown in the previous section SNMF can discover sparse features or independent components in data, even when they are not obviously clusterable. Therefore, it may be possible to use online SNMF as a generic signal processing tool.

More generally, online symmetric matrix factorization (SMF) is a powerful tool capable of performing multiple computations. In addition to clustering and feature discovery that online SNMF can perform, unconstrained SMF computes the principal subspace of the streamed data [17]. With the addition of a sparsity-inducing regularizer, SMF learns sparse dictionaries [30].

Given such versatility, online SMF may be a powerful and biologically plausible algorithmic model of neural computation. Therefore, we consider biological plausibility of the neural network implementation of the online SNMF next.

### B. Comparison with biological observations

The neural network implementation of the online SNMF algorithm we presented in this paper captures many features of real nervous systems.

Weighted summation of the input and non-negativity of neural output: Our online algorithm is implemented with units that perform a weighted summation of its inputs, followed by a rectification step. The non-negativity of our algorithm's output correctly reflects the monopolar nature of neuronal activity. Indeed, neuronal communication is implemented by action potentials or spikes, which rectify the membrane potential.

Local synaptic learning rules: Synaptic update rules in our network (8) depend only on pre- and postsynaptic activity and hence are biologically plausible. Feedforward synapses are updated according to the Hebbian rule, lateral synapses are updated according to the anti-Hebbian rule.

Output neurons obey Dale's law: Inspection of (8) reveals that while the feedforward synaptic weights in our network can be either positive or negative, depending on the sign of the input, lateral synaptic weights are always non-positive, due to the non-negativity of neural ouput. Therefore, the neurons in our network obey Dale's law.

Dependence of learning rate on cumulative activity: Synaptic weight update (8) predicts that synaptic plasticity decays with cumulative post-synaptic activity. Such variation of plasticity with time corresponds to the reports of long-term potentiation (LTP) decaying with age in an activity dependent manner [31]–[33].

Sparsity of neuronal activity: Our algorithm produces a neuronal firing distribution with a peak at zero and a heavy tail, Fig. 4B, in agreement with physiological measurements [24]. It is interesting that such distribution was achieved without including any sparsity-inducing regularizer into the SNMF cost funciton. Intuitively, the clustering operation of SNMF results in sparse firing by allowing only a few output neurons to be active for a given input.

## V. CONCLUSION

In this paper, we derive an online SNMF algorithm and demonstrate its performance in clustering and sparse feature discovery. We propose that the neural network implementation of the online SNMF of the similarity matrix may serve as the algorithmic theory of neural computation.

## ACKNOWLEDGMENTS

We would like to thank Sanjeev Arora, Alex Genkin, Eftychios Pnevmatikakis and Tao Hu for helpful discussions.